# Renormalized Multicanonical Sampling

David Yevick

Department of Physics

University of Waterloo

Waterloo, ON N2L 3G7

**Abstract:** For a homogeneous system divisible into identical, weakly interacting subsystems, the muticanonical procedure can be accelerated if it is first applied to determine of the density of states for a single subsystem. This result is then employed to approximate the state density of a subsystem with twice the size that forms the starting point of a new multicanonical iteration. Since this compound subsystem interacts less on average with its environment, iterating this sequence of steps rapidly generates the state density of the full system.

**Introduction:** Standard Monte-Carlo simulations model the density of states, here normalized to unity, as a function of energy, $p(\vec{E})$ - which coincides with the infinite temperature probability distribution function as obtained from an unbiased Monte-Carlo calculation - of a physical system quantified by one or more (typically macroscopic) variables $\vec{E}(\vec{\alpha})$ by assigning random values to the underlying (microscopic) parameters, $\vec{\alpha}$. While straightforward, physically significant rare events are inefficiently generated. Accordingly, the multicanonical [1] [2] [3] [4] [5] and Wang-Landau [6] [7] methods employ biased sampling to enhance the sampling probability of these events either computationally or experimentally [8]. To briefly summarize, specializing for simplicity to a single macroscopic variable $E$, small random changes in $\vec{\alpha}$ are accepted in accordance with a rule that favors displacements toward low probability regions of $\vec{E}$. The multicanonical method updates this rule iteratively starting from an initial Monte Carlo estimate, $p^{(1)}(E_i)$, of $p(E)$. Subsequently, an initial set of parameter values $\vec{\alpha}_{\text{current}}$ is selected and altered by a small, randomly generated quantity according to $\vec{\alpha}_{\text{new}} = \vec{\alpha}_{\text{current}} + \delta\vec{\alpha}$. This new realization is accepted with probability $\min\left(p^{(1)}(E_{\text{current}})/p^{(1)}(E_{\text{new}}),1\right)$ in which case $\alpha_{\text{current}}$ is replaced by $\alpha_{\text{new}}$. Repeating the sequence of displacement and acceptance steps is repeated a prescribed number of times yields a histogram $n(E_i)$ of the number of realizations recorded in each interval $i$ of $E$ (where $n(E_i)$ is also incremented after a rejected move from interval $i$ to $j$) that exhibits the desired bias towards lower probability regions. Multiplying $n(E_i)$ by $p^{(1)}(E_i)$, both of which are typically set to unity in unsampled histogram bins, generates a new probability distribution $p^{(2)}(E_i)$; the entire sequence of steps above are then iterated with $p^{(m)}(E)$ replaced by $p^{(m+1)}(E)$. As several iterations are required before the realizations sample the low probability configurations of interest, Wang and Landau multiplied the probability density in the *j*:th bin each time the bin is visited by a factor $\lambda^{(m)}$ that typically equals $ce^{1/m}$ for the $m$:th iteration of the method. This procedure converges to the desired result for large $m$.

**Numerical procedure:** In this paper, an alternative strategy is proposed for calculating $p(E)$ for large but effectively homogeneous systems that can be divided into weakly coupled subsystems. To illustrate, if a linear homogeneous spin system containing $N = \Lambda \cdot 2^M$ spins with short-range interactions is divided into subsystems containing $\Lambda \cdot 2^K$ spins with $K < M$, the density of states $p_{S^{(K)}}(E)$ of a single subsystem is rapidly obtained for any choice of boundary conditions. Since the interaction between sites is short-range, the effective interaction between two such subsystems is smaller than the individual spin-spin coupling. Hence the probability that a combination of two adjacent identical subsystems $S_1 + S_2$ occupies a certain state $\{\{\vec{\alpha}_1\},\{\vec{\alpha}_2\}\}$ with a total value $E = E_1 + E_2$ can be approximated by the product of the probabilities that the isolated systems $S_1$ and $S_2$ occupy the states $\{\vec{\alpha}_1\}$ and $\{\vec{\alpha}_2\}$. The probability of a realization with system variable $E$ in the combined system can therefore be estimated by the convolution $p_{(S_1+S_2)^{(K+1)}}(E) = \int_0^{E_{\max}} p_{S_1^{(K)}}(E_1) p_{S_2^{(K)}}(E - E_1) dE_1$, which can subsequently be improved by one or more multicanonical iterations. As this process is iterated, the effective interaction between successively larger subsystems monotonically decreases, enhancing the accuracy of the convolutions.

While more numerically efficient than the multicanonical or Wang-Landau methods, the implementation of the renormalized procedure requires additional code for the convolution. As well, the multicanonical calculations must sample every value of $E$ in the computational window of the subsystem while the transition rule must ensure that transitions occur sufficiently frequently out of states with very low probabilities at the boundaries of the computational window.

**Computer program:** The algorithm of this paper in its simplest form can be immediately incorporated into a standard multicanonical program such as that appearing on page 225 of [9] by inclusion of the first and last five lines of the following Octave/MATLAB code, where the syntax (e.g. spacing and naming) conventions are those of [9] [10]. The multicanonical procedure is inserted without modification as indicated into the remainder of the program. Optimization of the variables **numberOfConvolutions** and **numberOfRealizations** requires additional lines of code but can also be implemented manually by viewing a graph of $p(E)$ after each multicanonical iteration.

```
histogramR = ones( 1, numberOfSteps + 1 );
for loopConvolution = 1 : numberOfConvolutions + 1
    % Beginning of standard multicanonical code
    for loopOuter = 1 : numberOfMulticanonicalLoops;
        histogramNewR = ones( 1, numberOfSteps + 1 );
        % ...
        for loop = 1 : numberOfRealizations;
        % ...
        end
        histogramR = histogramR .* histogramNewR;
    end
    % End of standard multicanonical code ...
```

```
      if ( loopConvolution == numberOfConvolutions + 1 ) break; end;
      histogramR = histogramR / sum( histogramR );
      numberOfSteps = 2 * numberOfSteps;
      histogramR = conv( histogramR, histogramR );
end
```

When applying the algorithm to e.g. the one-dimensional Ising model its efficiency and accuracy can be significantly increased by modifying the multicanonical acceptance rule to read, where **upperHistogramIndex** corresponds to the largest possible value of $E$,

```
if ( rand < histOld / histNew | histogramIndexNew == histogramIndexOld ...
        | histogramIndexOld == 1 | histogramIndexOld == upperHistogramIndex )
% transition accepted
end
```

This facilitates the sampling of multiple states in low-probability regions of $E$ in which most transitions are rejected (since transitions can still occur between realizations with identical $E$ values) while avoiding possible difficulties associated with boundary effects. The behavior at the boundary points can then be approximated by including the following lines

```
histogramR(1) = histogramR(2)^2 / histogramR(3);
histogramR(upperHistogramIndex) = histogramR(upperHistogramIndex - 1)^2 / ...
      histogramR(upperHistogramIndex - 2);
```

immediately after the statement

```
histogramR = histogramR .* histogramNewR;
```

in the above program.

**Numerical results:** To verify the accuracy of the "renormalized" multicanonical procedure, a trivial system of noninteracting components is first considered. In particular, the probability distribution function of heads for **numberOfSteps =** 160 coin flips is analyzed starting with **numberOfSteps =** 10 coins with a renormalized multicanonical calculation employing **numberOfMulticanonicalLoops** = 3 multicanonical iterations of **numberOfRealizations =** 80,000 realizations preceding each renormalization step the subsystem size is doubled until **numberOfConvolutions** = 4. The result of this procedure (dashed line) is compared to the result of a standard multicanonical calculation with twenty 80,000 realization iterations (+ markers superimposed on dashed-dotted line) and the exact result (solid line) in Fig. 1. The accuracy of the standard multicanonical and renormalized multicanonical methods are both principally governed by the number of realizations for each multicanonical iteration and are therefore nearly identical. However, more generally the increase in range for a given multicanonical iteration in the standard method gradually decreases with iteration number, unlike the renormalized procedure for which the range expands rapidly as each additional convolution is applied.

The renormalized multicanonical results for the normalized density of states $p(E)$ of the one-dimensional Ising model with unit coupling constant and half-integral spins for zero external magnetic field is next displayed in Fig. (2) where the rescaled energy, $E$, is taken to be the sum of half the interaction energy and ¼ times the total number of spins (so that the smallest value of $E$ always coincides with the origin while the possible values $E$ are unit spaced integer values for an even number of spins).  The calculation is performed with periodic boundary conditions, **numberOfMulticanonicalLoops =** 8, **numberOfRealizations** = 400,000, **numberOfConvolutions** = 2 for 16 initial spins (and hence 9 discrete energy states) and an antiferromagnetic interaction.  The graph displays the results obtained after one, two and eight multicanonical refinement steps at each stage of the calculation where the artificial boundary condition causes $p(E)$ at the 3 points at each computational window boundary to vary linearly with $E$.  The stability of the procedure is evident from the number of events recorded for each value of $E$, c.f. Fig. (3) which demonstrate that after each convolution only a single multicanonical iteration is required before the number of samples in each bin becomes nearly independent of bin number.

**Discussion and conclusions:**  The numerical algorithm introduced above is conceptually related to the renormalization group procedure in that a property of a uniform system composed of identical units is calculated by computing its value for successively larger and more weakly interacting identical subunits.   In contrast to competing techniques, the dynamic range and accuracy therefore increase rapidly as the calculation progresses.  Further, the method can be immediately applied to multidimensional systems described by a single macroscopic parameter or, by employing multidimensional convolutions, to multiple macroscopic parameters, $\vec{E}$.  However, to implement a version of the algorithm that does not require user interaction requires additional lines of code that ensure proper sampling of the computational window is properly sampled before each doubling of the system size.  Additionally, as increasingly smaller probabilities are evaluated the accuracy of the method cannot exceed that of the underlying multicanonical calculations, which can however be improved through well-documented techniques. [11] [12]

**Acknowledgements:**  The Natural Sciences and Engineering Research Council of Canada (NSERC) is acknowledged for financial support.

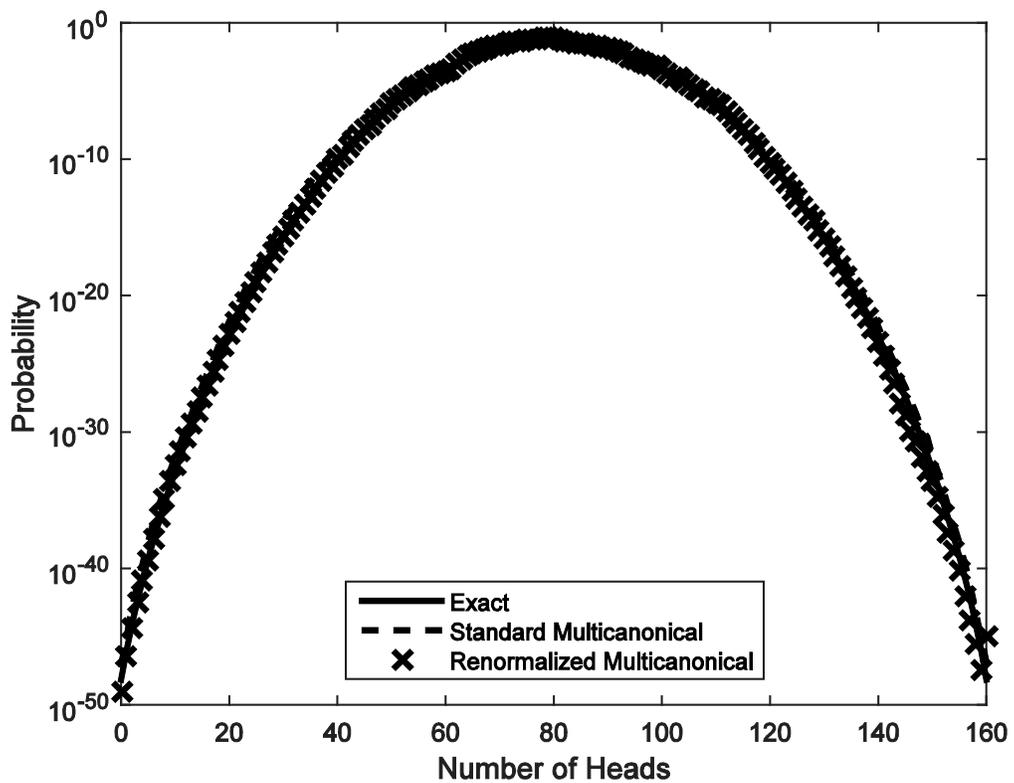

*Figure 1: The probability distribution function of heads for **numberOfSteps** = 160 coin flips as calculated with a four stage renormalized multicanonical evaluation (x), the standard multicanonical method (dashed line) and the exact result (solid line).*

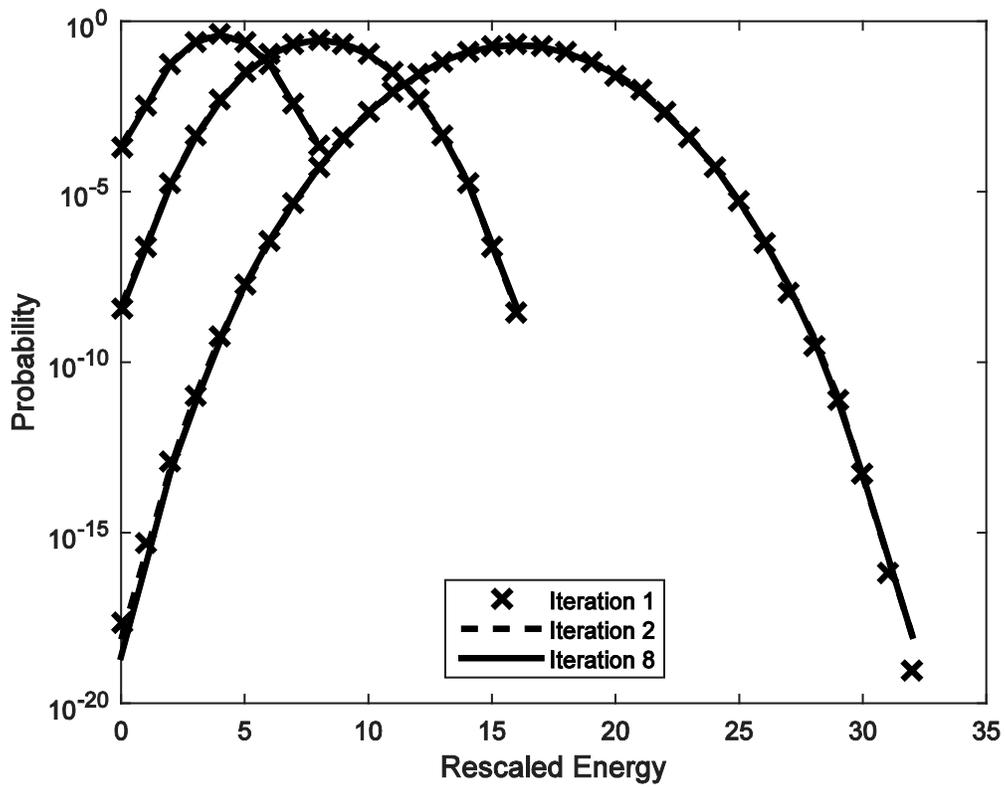

Figure 2: The renormalized multicanonical results after 1, 2 and 3 iterations:

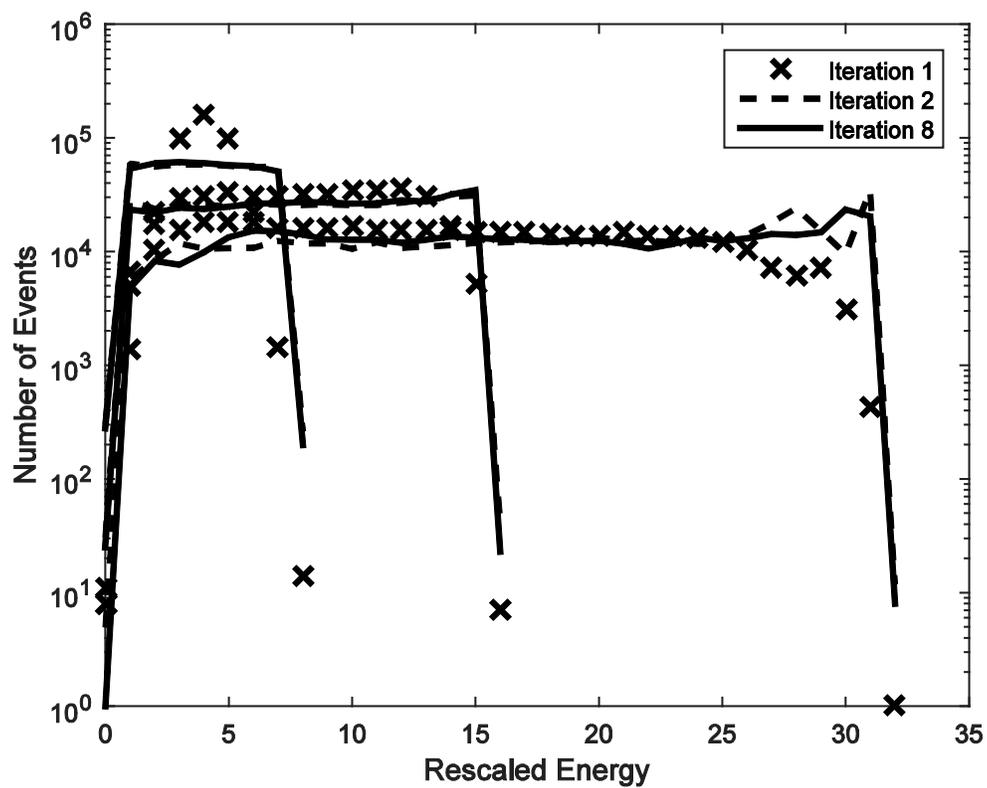

Figure 3: The intermediate multicanonical histograms generated during the 3 iterations